\documentclass[review]{elsarticle}
\usepackage{color}
\usepackage{hyperref}

\journal{Journal of \LaTeX\ Templates}
\usepackage{natbib}

\begin{document}

\begin{frontmatter}

\title{Implications of Neutron Star Properties for the Existence of Light Dark Matter}

\author{T.~F.~Motta}
\address{CSSM and ARC Centre of Excellence for Particle Physics at the Terascale, Department of Physics, University of Adelaide SA 5005 Australia}
\author{P.~A.~M.~Guichon}
\address{IRFU-CEA, Universit\'e Paris-Saclay, F91191 Gif sur Yvette, France}
\author{A.~W.~Thomas}
\address{CSSM and ARC Centre of Excellence for Particle Physics at the Terascale, Department of Physics, University of Adelaide SA 5005 Australia}

\begin{abstract}
It was recently suggested that the discrepancy between two methods of measuring the lifetime of the neutron may be a result of an unseen decay mode into a dark matter particle which is almost degenerate with the neutron. We explore the consequences of this for the properties of neutron stars, finding that their known properties are in conflict with the existence of such a particle.
\end{abstract}


\end{frontmatter}


\section{Introduction}

For some time there has been a discrepancy of about 8 seconds ( $3.5 \sigma$)between two techniques used to determine the lifetime of a free neutron. Following earlier suggestions that this discrepancy might result from an oscillation to a mirror neutron~\cite{Serebrov:2007gw}, it was recently proposed by Fornal and Grinstein~\cite{Fornal:2018eol} that it could rather be caused by a new decay mode of the neutron to an almost degenerate dark matter particle, which would not be visible in one of the measurments. Several authors have already placed further constraints on this mechanism. Czarnecki and collaborators~\cite{Czarnecki:2018okw} noted a degree of tension between the current best value of the neutron axial charge and this explanation but could not rule it out. On the basis of a new experimental measurement at the Los Alamos ultra-cold neutron facility, Tang {\it et al.}~\cite{Tang:2018eln} ruled out the decay mode $n \rightarrow DM + \gamma$, where $DM$ represents the hypothesised dark matter particle, however this leaves the decay to two dark matter particles as a viable possibility. More recently it has been suggested by 
Serebrov {\it et al.}~\cite{Serebrov:2018mva} that a so-called reactor anti-neutrino anomaly might also be explained by the same mechanism.

Here we examine the consequences of the existence of such a dark matter particle in a very different environment, namely a neutron star. If there were indeed a dark matter particle almost degenerate with the neutron, then as the density of nuclear matter increases and the chemical potential of the neutrons rises above the mass of the dark matter particle the neutrons must decay to restore chemical equilibrium. Since the composition of the nuclear matter in the core of a neutron star is dominated by neutrons with a chemical potential well above the neutron mass, in this new scenario one must expect that almost half of the energy density of matter in the core would now be dark. As we shall see this dramatically reduces the pressure for a given energy density, which in turn reduces the maximum mass of the neutron stars which can be formed. Indeed, our calculations suggest that this scenario is incompatible with the existence of neutron stars whose masses are already well established.

\section{Equations of state and solution of the TOV equations}

In order to calculate the equation of state (EoS) of dense matter, which is required as input to the Tolman-Oppenheimer-Volkov (TOV) equations~\cite{Tolman:1939jz,Oppenheimer:1939ne} used to compute the mass and radius of a given neutron star, we use the quark-meson coupling (QMC) model~\cite{RikovskaStone:2006ta,Guichon:1987jp,Guichon:1995ue}. This model yields a relativistic EoS for hadrons starting from the self-consistent solution for the structure of hadrons moving in relativistic mean-fields corresponding to the $\sigma , \omega$ and $\rho$ mesons. It naturally yields three-body forces between hadrons with no additional parameters~\cite{Guichon:2004xg} and as a consequence can support neutron stars with masses in the region of two M$_\odot$, {\em even when} hyperons are 
included~\cite{RikovskaStone:2006ta}, while producing realistic hypernuclear binding 
energies~\cite{Guichon:2008zz,Saito:2005rv}. 
\begin{figure}
 \centering
 \includegraphics[scale=0.7]{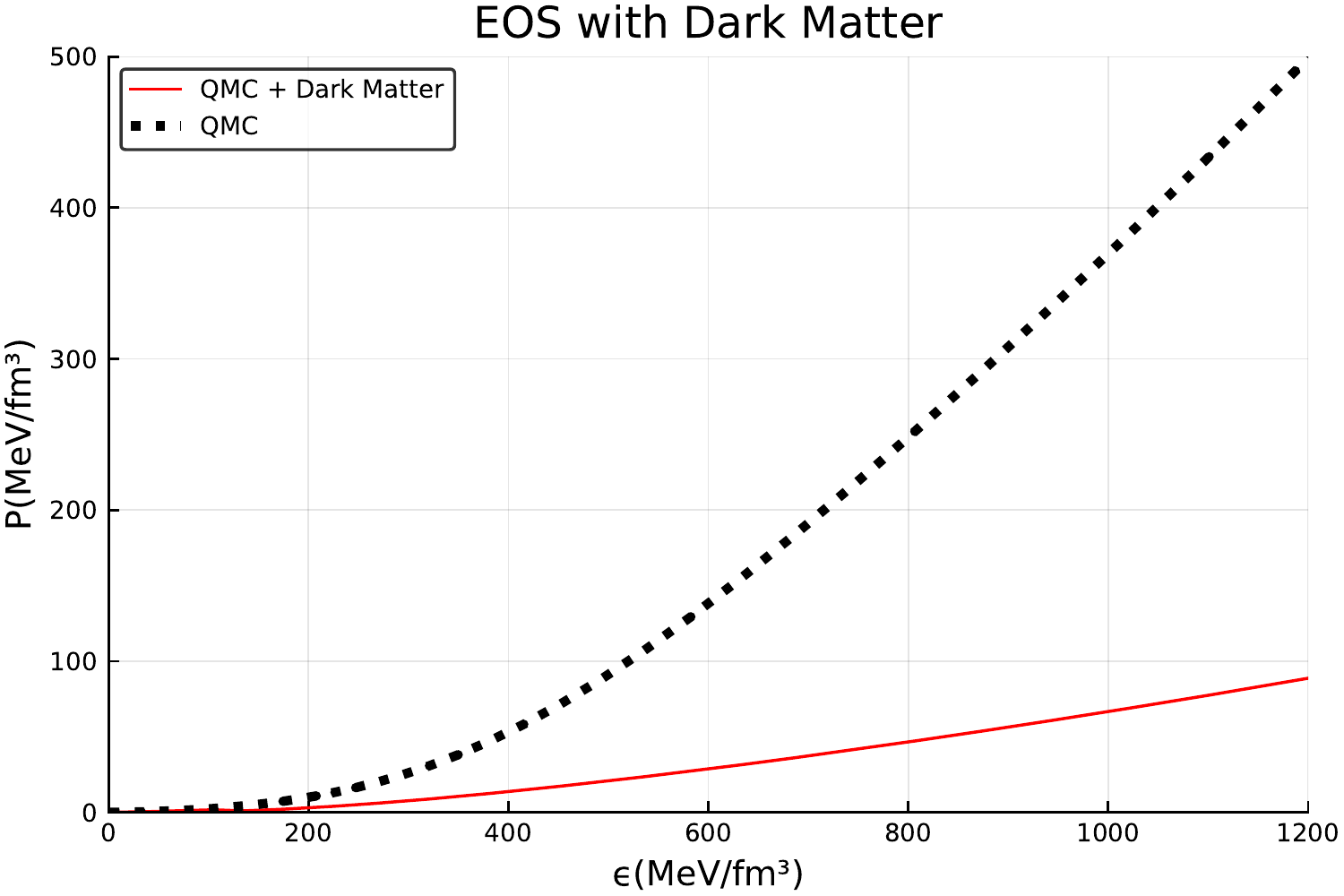}
  \caption{Comparison of the equation of state (EoS) for nuclear matter in $\beta$-equilibrium, with and without the inclusion of a dark matter particle degenerate with the neutron. Note the dramatic softening of the EoS. }
  \label{fig:EoS}
\end{figure}

On the other hand, for the present application many of these features are not needed, as so much of the energy density is carried by dark matter that the maximum stellar mass is already reached before hyperons can enter the EoS, so it enough to consider matter consisting of neutrons, protons, electrons and muons in $\beta$-equilibrium, in chemical equilibrium with dark matter. In this case, treating the DM as degenerate with the neutron (as a difference of order 1 MeV will make no meaningful difference), the chemical equilibrium equations are simply
\begin{equation}
\mu_n = \mu_p + \mu_e \,\,\, , \,\,\, \mu_n=\mu_DM \, ,
\label{eq:chemical}
\end{equation}
where $\mu_{n(p)}$ are the proton and neutron chemical potentials, $\mu_e$ is the electron chemical potential and we also include the muon if it is energetically allowed. 
\begin{figure}
 \centering
\includegraphics[scale=0.7]{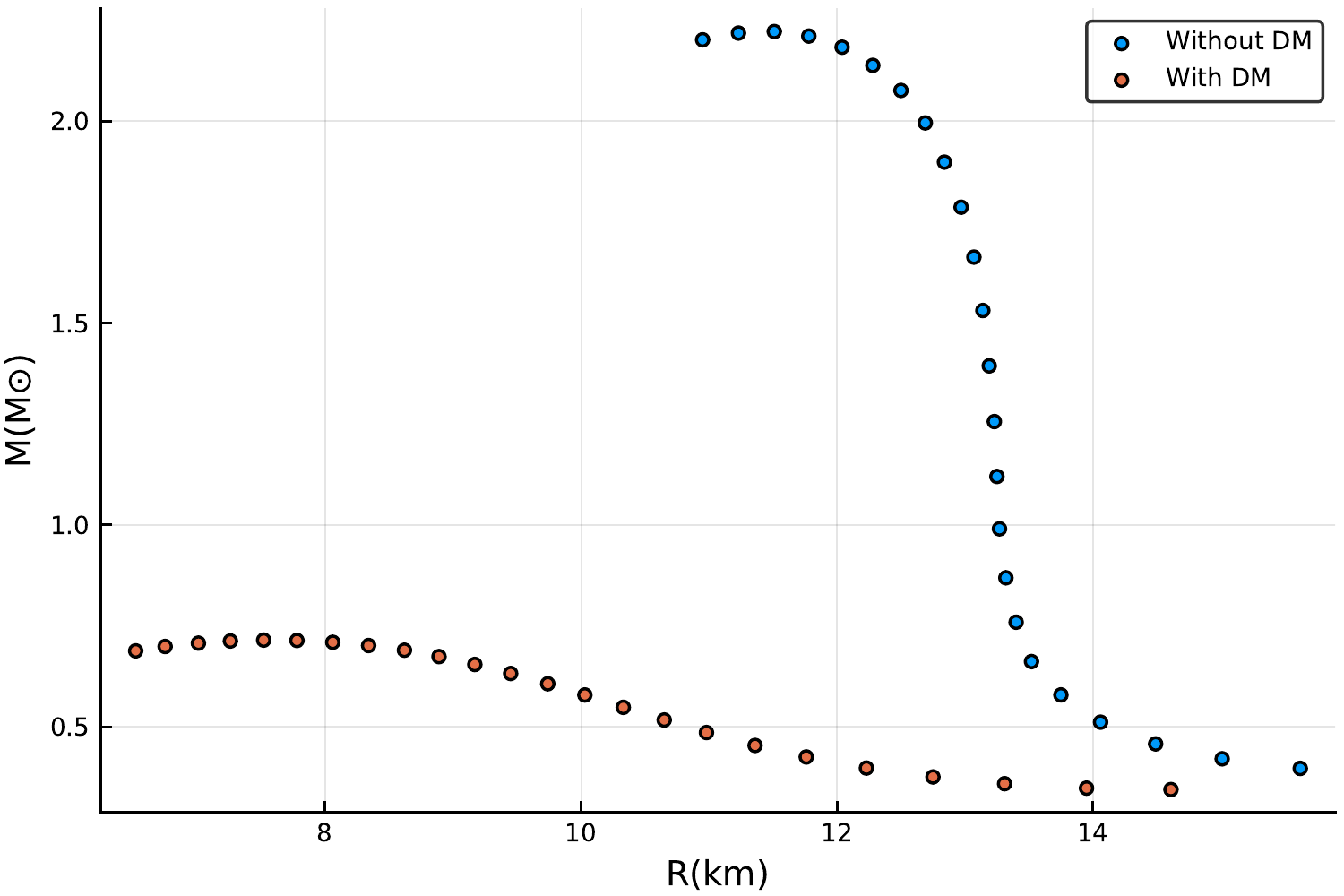}
  \caption{Comparison of the mass versus radius curves resulting from the solution of the TOV equations for the case where only nucleons are allowed and where dark matter is included. The reduction in the maximum possible mass is dramatic, from around 2.2 M$_\odot$ to just 0.7 M$_\odot$.  }
  \label{fig:massvsR}
  \end{figure}
\begin{figure}
 \centering
 \includegraphics[scale=0.7]{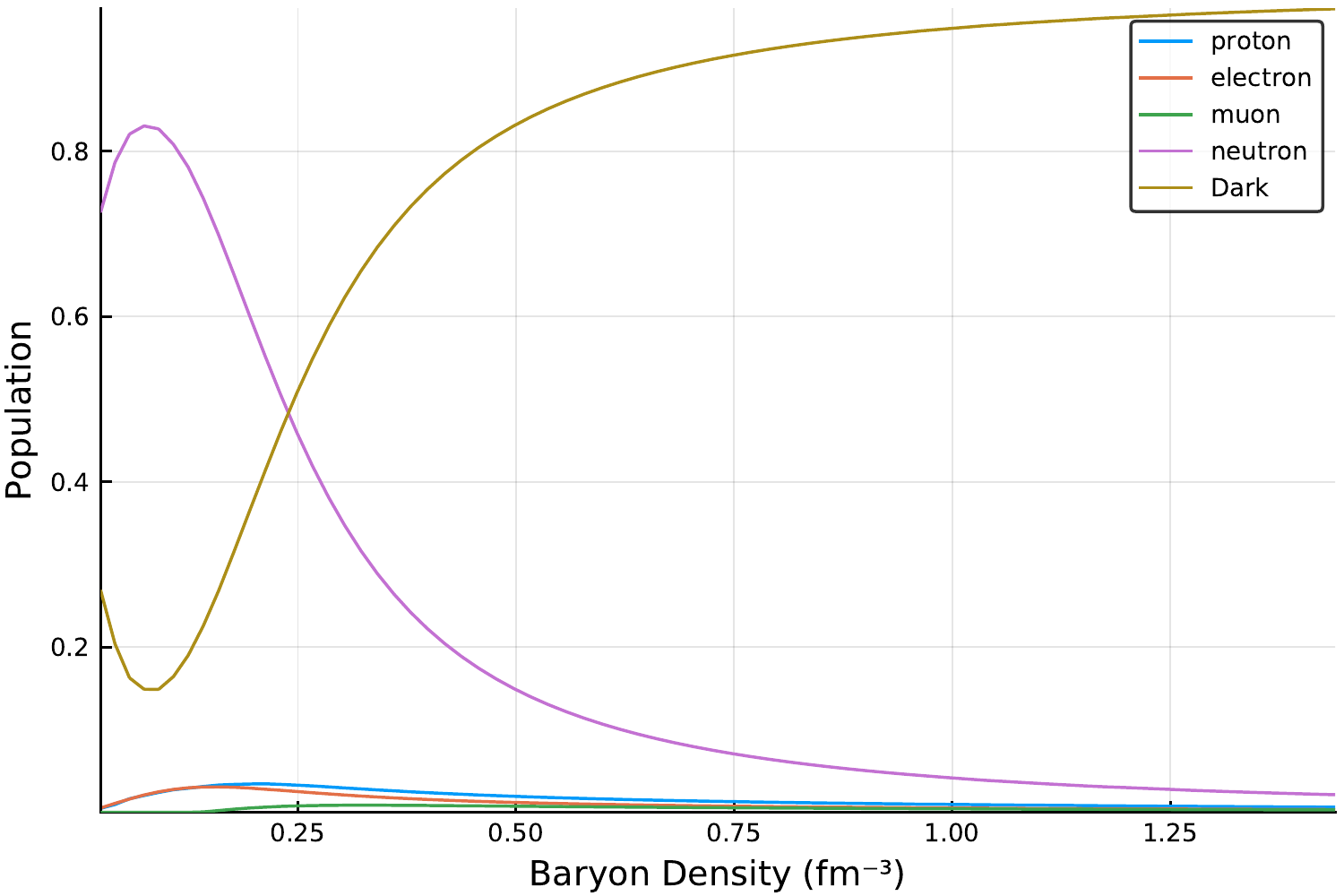}
  \caption{Illustration of the species fractions appearing in matter in $\beta$ and chemical equilibrium when dark matter is included, as a function of baryon density. Clearly dark matter begins to dominate beyond 2 $\rho_0$. }
  \label{fig:species}
  \end{figure}

To calculate the EoS we use the model QMC700, which is fully explained in Ref.~\cite{RikovskaStone:2006ta}. In the case where only nucleons are included, like most other relativistic mean field treatments this yields a maximum neutron star mass around 2.2 M$_\odot$. The expression for the pressure is
\begin{equation}
P = \sum_i \mu_i n_i - \epsilon \, ,
\label{eq:pressure}
\end{equation}
where $n_i$ are number densities of the species present, $\mu_i$ their chemical potentials and $\epsilon$ the total energy density. From Eq.~(\ref{eq:pressure}) it is obvious that at the same energy density the lower neutron chemical potential resulting when DM is included will result in a lower pressure. While the sign of the effect is obvious, the size of the reduction in practice is remarkable, as we see in Fig.~\ref{fig:EoS}. 

Such a reduction in pressure for a given energy density might be expected to lead to stars with a lower maximum mass but the extent of the reduction is very large. In particular, as illustrated in Fig.~\ref{fig:massvsR}, the maximum mass of a neutron star decreases from around 2.2 M$_\odot$ to as low as 0.7 M$_\odot$ once DM is allowed. 
In part this can be understood by the argument presented earlier, below Eq.~(\ref{eq:pressure}). However, the effect is significantly enhanced by the change in composition shown in Fig.~\ref{fig:species}, where we see that the composition of the matter in the core of the star is dominated by DM at relatively low baryon density.

\section{Concluding remarks}

In concluding, it seems worthwhile to put our results in some perspective. Many authors have investigated the consequences for neutron stars if they capture dark matter of various kinds~\cite{ PerezGarcia:2011hh, Perez-Garcia:2014dra, Li:2012ii, PerezGarcia:2010ap}. The situation here is very different. Under the scenario proposed by Fornal and Grinstein the lifetime of a neutron which is forbidden by energy conservation and the Pauli principle to $\beta$-decay will be of the order of days. Thus in a supernova explosion a proto-neutron star would form and then decay over a period of days and weeks, as the neutrons high in the Fermi sea are converted to dark matter. 

Most of the neutron stars which have been observed have masses around 1.4-.15 M$_\odot$~\cite{Kiziltan:2013oja}, while the maximum mass neutron stars that have been discovered have masses around 2 M$_\odot$~\cite{Demorest:2010bx,Antoniadis:2013pzd}, corresponding to a central density of around 5-6 times nuclear matter density~\cite{ RikovskaStone:2006ta,Whittenbury:2013wma}. The latter corresponds to a total baryon number of order $3.2 \times 10^{57}$. It seems reasonable that the almost degenerate dark matter particles formed when the neutrons decay would remain trapped by gravity while their lighter decay partners escape, in which case the total number of dark matter particles plus nucleons after the decay process would be approximately equal to the total number of nucleons before decay. Such a star is well beyond the maximum mass consistent with stability against collapse and would become a black hole. Indeed, the number of dark matter particles plus nucleons in the dark matter star with the maximum mass is just $9.2 \times 10^{56}$. This corresponds to a star, before decay, of mass around 0.7 M$_\odot$.

We have explored the consequences for the maximum mass of a neutron star of the proposed explanation for the discrepancy between modern measurements of the lifetime of the neutron involving a decay mode to a dark matter particle which is almost degenerate with the neutron. The dark matter is assumed to be non-interacting. As the central density of the star increases it is unavoidable that the fraction of particles that are dark begins to dominate, with the consequence that for a given energy density the pressure is dramatically lowered. As a consequence, when the TOV equations are solved the pressure is unable to sustain stars as massive as those found without dark matter. Indeed the maximum mass allowed is reduced to just 0.7 M$_\odot$. 

We have verified that alternate parameter sets with or without the inclusion of hyperons make no significant difference. The effect of this hypothesised dark matter particle is so dramatic that the conclusion that one cannot generate stable stars with the masses observed in Nature, typically around 1.4 M$_\odot$ but as large as 2 M$_\odot$, is model independent.  As a result we are led to conclude that this explanation of the discrepancy in neutron lifetimes cannot be correct and the proposed dark matter particle does not exist.

\section*{Acknowledgements}
This work was supported by the University of Adelaide and by the Australian Research Council through the 
ARC Centre of Excellence for Particle Physics at the Terascale (CE110001104) and Discovery Project DP150103164.

\section*{References}


\end{document}